\documentclass[aps, preprint, nofootinbib, superscriptaddress, showpacs, longbibliography]{revtex4-1}
\usepackage{graphicx}
\usepackage{color}
\usepackage[colorlinks=false]{hyperref}
\usepackage[textsize=footnotesize]{todonotes}
\usepackage[todos]{CMImacros}
\usepackage{lineno}
\usepackage{mathrsfs}
\usepackage{amsmath}
\usepackage{amssymb}
\usepackage{braket}
\usepackage[utf8]{inputenc}
\date{\today}

\usepackage{color}
\definecolor{oldtxtcolor}{rgb}{0.00, 0.0, 0.5}
\definecolor{newtxtcolor}{rgb}{0.00, 0.3867, 0.00}
\definecolor{newtxtcolor}{rgb}{0.00, 0.0, 1}
\definecolor{oldtxtcolor}{rgb}{1.00, 0.0, 0.00}

\def\verX{12}
\def\verO{1}
\def\verN{2}
\def\verON{12}

\ifx\verX\verO
 \newcommand { \oldtxt }[1] {{\color{oldtxtcolor}{#1}}}
 \newcommand { \newtxt }[1] {}
\fi
\ifx\verX\verN
 \newcommand { \oldtxt }[1] {}
 \newcommand { \newtxt }[1] {{\color{newtxtcolor}{#1}}}
\fi
\ifx\verX\verON
 \newcommand { \oldtxt }[1] {{\color{oldtxtcolor}{#1}}}
 \newcommand { \newtxt }[1] {{\color{newtxtcolor}{#1}}}
\fi

\begin{document}

\title{Perspective: Ultrafast Imaging of Molecular Dynamics Using Ultrafast Low-Frequency Lasers, X-ray Free Electron Laser and Electron Pulses}
\author{Ming Zhang}
\affiliation{State Key Laboratory for Mesoscopic Physics and Collaborative Innovation Center of Quantum Matter, School of Physics, Peking University, Beijing 100871, China}
\author{Zhengning Guo}
\affiliation{State Key Laboratory for Mesoscopic Physics and Collaborative Innovation Center of Quantum Matter, School of Physics, Peking University, Beijing 100871, China}
\author{Xiaoyu Mi}
\affiliation{State Key Laboratory for Mesoscopic Physics and Collaborative Innovation Center of Quantum Matter, School of Physics, Peking University, Beijing 100871, China}
\author{Zheng Li}
\email{zheng.li@pku.edu.cn}
\affiliation{State Key Laboratory for Mesoscopic Physics and Collaborative Innovation Center of Quantum Matter, School of Physics, Peking University, Beijing 100871, China}
\affiliation{Collaborative Innovation Center of Extreme Optics, Shanxi University, Taiyuan, Shanxi 030006, China}
\affiliation{Peking University Yangtze Delta Institute of Optoelectronics, Nantong, China}
\author{Yunquan Liu}
\email{yunquan.liu@pku.edu.cn}
\affiliation{State Key Laboratory for Mesoscopic Physics and Collaborative Innovation Center of Quantum Matter, School of Physics, Peking University, Beijing 100871, China}
\affiliation{Collaborative Innovation Center of Extreme Optics, Shanxi University, Taiyuan, Shanxi 030006, China}
\affiliation{Center for Applied Physics and Technology, HEDPS, Peking University, Beijing 100871, China}
\maketitle
\newpage

\begin{center}
    \large{Abstract}
\end{center}

The requirement of high space-time resolution and brightness is a great challenge for imaging atomic motion and making molecular movies. Important breakthroughs in ultrabright tabletop laser, x-ray and electron sources have enabled the direct imaging of evolving molecular structures in chemical processes.
And recent experimental advances in preparing ultrafast laser and electron pulses equipped molecular imaging with femtosecond time resolution. 
This Perspectives present an overview of versatile imaging methods of molecular dynamics.
{High-order harmonic generation imaging and photoelectron diffraction imaging are based on laser-induced ionization and rescattering processes.}
Coulomb explosion imaging retrieves molecular structural information by detecting the momentum vectors of fragmented ions. 
Diffraction imaging encodes molecular structural and electronic information in reciprocal space. 
We also present various applications of these ultrafast imaging methods in resolving laser-induced nuclear and electronic dynamics.
\begin{figure}
    \centering
    \includegraphics[width=2in]{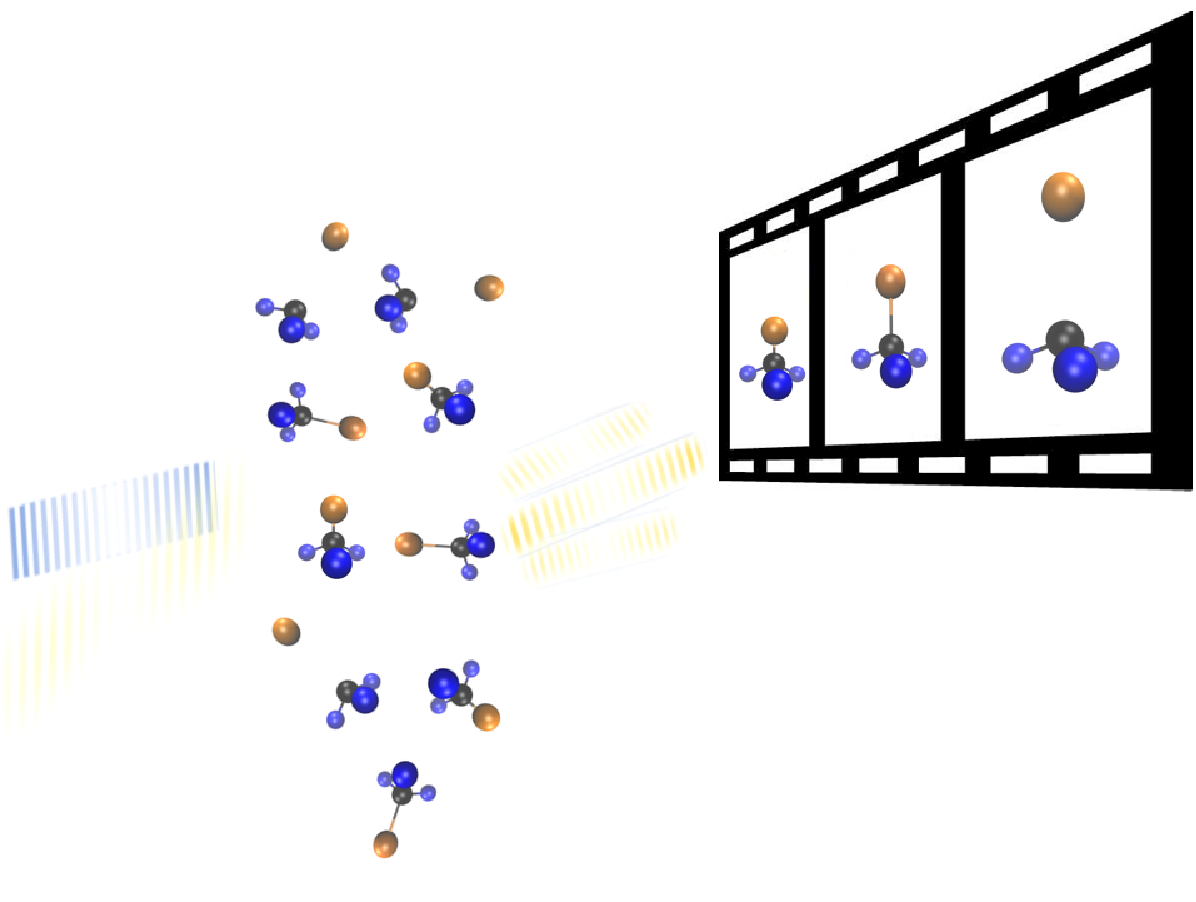}
    \caption{TOC image.}
\end{figure}

\newpage
\begin{figure}
    \centering
    \includegraphics[width=16cm]{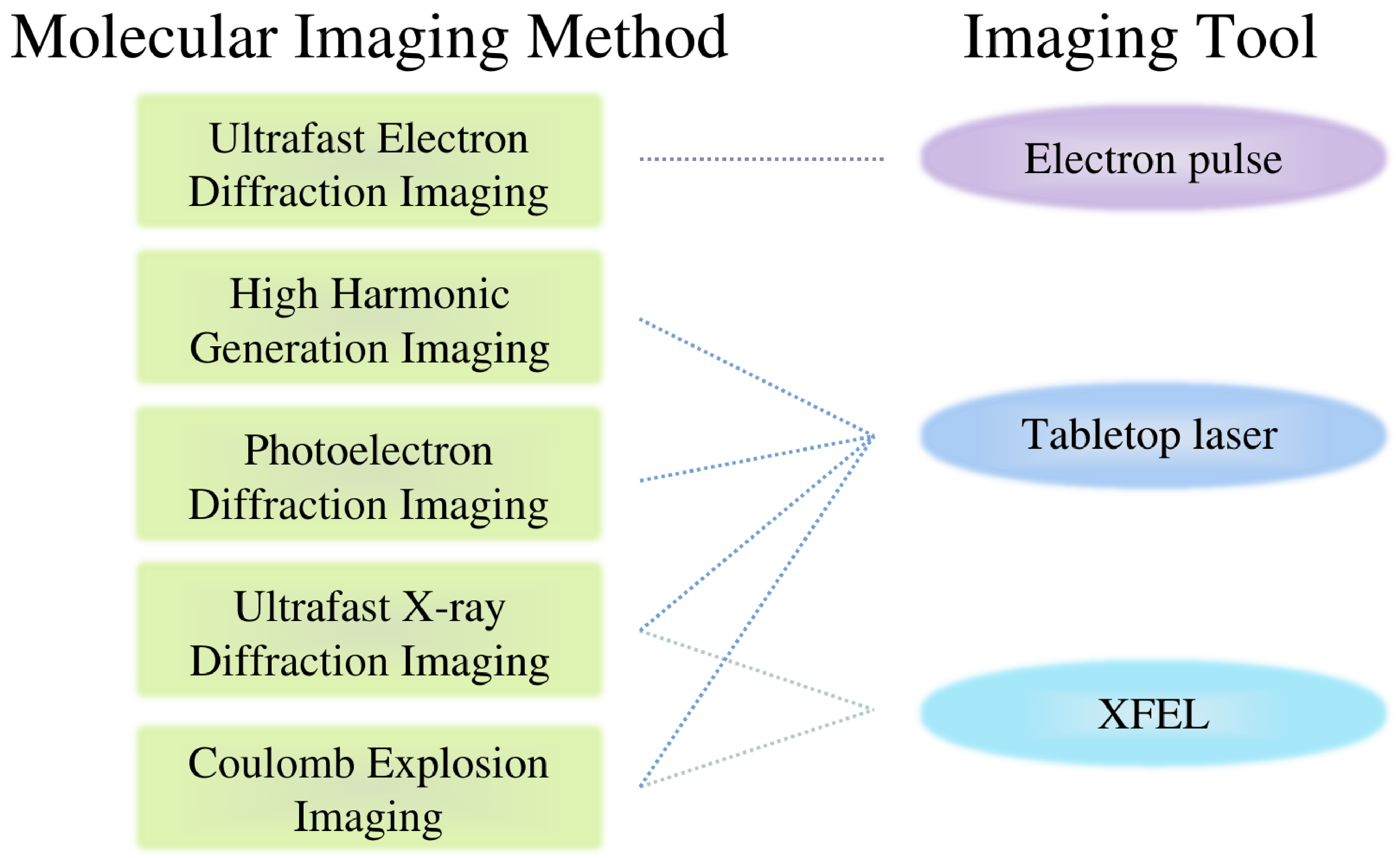}
    \caption{\label{fig:roadmap}
    The state of the art methods and experimental tools for ultrafast imaging of molecular dynamics using tabletop laser, x-ray free electron laser (XFEL) light sources and ultrashort electron pulses.}
\end{figure}

 The imaging of complex molecular structures with electron or x-ray diffraction is challenging and it is especially difficult to extract dynamic structural information from their diffraction patterns with femtosecond time resolution, which is the characteristic time scale of atomic motion in molecules. Direct observation of atomic motions in real time during chemical reactions is thought as a Gedanken experiment for chemists ~\cite{Polanyi95:ACR119,Miller14:ARPC583}. 
Given the extremely high spatial and temporal resolution required to image atomic motions along reaction coordinates, it was thought to be the purest form of a Gedanken experiment. %
It was even argued by Manfred Eigen in his Nobel lecture~\cite{Eigen67} that we could never achieve the necessary imaging technologies to see atoms rearrange in space during
chemical reactions, as he argued that we would always have to use
indirect methods for inferring the underlying mechanism of chemical reactions.
Despite the challenge for high spatio-temporal resolution requirement, it has been a pedagogical tool for theoretical conceptualization of chemical reactions by transition state involving molecular structural dynamics at the barrier-crossing regions, and recent experimental advances in ultrabright tabletop laser, x-ray and electron sources have brought the Gedanken experiment of directly imaging molecular motion into reality~\cite{Ischenko17:CR11066}. 

 Considering the spatial resolution for experimentally imaging such structural dynamics, the length scale for chemical bonds is \AA{}ngstrom and the accurate value can be determined from static gas phase electron diffraction with 0.001 \AA~precision~\cite{Ischenko17:CR11066,Ischenko19:054201}. For example, {the transient hydrogen bond contraction of $\sim$0.04~\AA~in liquid water is measured by ultrafast electron diffraction~\cite{YangJ21:Nature596}.} In chemical reactions, intramolecular and intermolecular electron transfer {changes of the potential energy for the system, and leads to changes of nuclear configuration for stabilization of system energy~\cite{Marcus97:JEC251,Kosower86:ARPC127}.  For example, photoexcitation of ammonium sulfate~\cite{Woerner10:064509} induces a sub-100 fs electron transfer from the sulfate groups into a highly confined electron channel, and the geometry is stabilized by transferring protons.} The magnitude order of molecular structural variation in electron transfer processes is 0.01 \AA~to 0.1 \AA~\cite{Kubar08:JPB8788,Oberhofer09:JCP064101}, which is feasible to be imaged by high energy electrons. {Ultrafast x-ray powder diffraction observed the change of distance between oxygen atoms of adjacent SO$_4^{2-}$ groups in ammonium sulfate of $\sim$0.2~\AA~amplitude~\cite{Woerner10:064509}.} 
Apart from the spatial resolution requirements, achieving the femtosecond time resolution is also an essential factor in catching the very moment in chemical reaction. In unimolecular reactions, time scale corresponding to the thermal sampling frequency of a barrier between reactant and product is 100 fs at room temperature. These frequencies in thermal sampling involve motions damped by anharmonic coupling in reaction coordinates. The time scale for intramolecular vibrational energy redistribution is also 100 fs, which is viewed as the shutter speed for imaging the molecular dynamics in reaction modes~\cite{Dwyer06:PTRSA741,Sciaini11:RPR096101}. For chemical process involving strongly repulsive excited state potentials like photodissociation~\cite{Stankus16:FD525}, the time scale for {dissociation coordinates} is 10 fs. But for most reactions, the time scale for {time resolution for most chemical processes} is 100 fs to ps, for example, photoisomerization of retinal~\cite{Johnson14:PCCP21310}, electron transfer~\cite{Marcus85:BBARB265,Mark89:Science1674}, proton transfer~\cite{Paul92:Science975} and ring-closing reaction~\cite{Jean13:JPCB117}. Thus we regard 10-100 fs as the time scale of chemistry.

The challenge to image chemical process can be generalized as achieving the {requirement of average beam current} of electron and {average flux of x-ray pulses} with sub-\AA~spatial resolution and subpicosecond temporal resolution.
For more than a hundred years, sub-\AA~spatial resolution has been reached by using x-rays with sufficiently short wavelengths to determine atomic positions. Over 20 years ago, plasma generation or similar electron pulse generation techniques via photoemission gave birth to the 100 fs pulses and the subpicosecond time resolution was reached. But these techniques are unable to make femtosecond movies directly in molecules because of their low brightness, and stroboscopic pump-probe method was applied to enhance the total brightness~\cite{Lobastov05:PNAS012}.{In solid state systems, because stroboscopic measurement requires the system to recover from pump and probe pulse~\cite{Chapman09:Natmat8}, only reversible processes such as phonon excitation~\cite{Bargheer04:Science306} can be detected nondestructively.} But the structural change and excitation dynamics make the sampling process irreversible, which limits the application scope for these techniques. {However, for gas phase diffraction, the sample is refreshed after every shot and the dynamics do not need to be reversible. }

With the advance of ultrafast high intensity laser technology, strong-field laser induced tunneling, laser-induced electron diffraction, high harmonic generation imaging and Coulomb explosion imaging have been introduced as new tools into the field of ultrafast imaging of structural dynamics. The imaging methods using high-order harmonic generation (HHG) and photoelectron diffraction are used for reconstructing nuclear dynamics and tomographic imaging of molecular Dyson orbitals. Using the relationship between HHG signal and the nuclear autocorrelation function~\cite{Lein05:PRL94}, the nuclear motion of H$_2^+$ and D$_2^+$ can be reconstructed, and attosecond temporal resolution is reached through mapping between the harmonic order of photons to time delay of electron trajectory~\cite{Lan17:PRL119,baker06:424}. Beyond nuclear degrees of freedom, the highest occupied molecular orbital (HOMO) of molecules, which is a good approximation of the Dyson orbital in outer valence photoionization, can be selectively and tomographically reconstructed by recording HHG spectra for different alignment angles under single active electron approximation~\cite{Ville04:Nature432,Patch07:JCP126}. Apart from HHG imaging, the tunneling ionized photoelectron rescattering off molecular ion also provides a self-imaging method, laser-induced electron diffraction (LIED)~\cite{Pullen16:11922}. {For example, the electronic orbital and nuclear position of N$_2$ and O$_2$ can be extracted by the momentum distribution of scattered electrons from LIED~\cite{Meckel08:Science320}.} The dissociation dynamics of di-ionized acetylene (C$_2$H$_2^{2+}$) can be visualized by extracting the temporal evolution of bond distances from LIED experimental data~\cite{wolter16:Science308}, and Dyson orbital of ionized molecules is incorporated in the photoelectron angular distribution (PAD)~\cite{Oana07:JCP127}. 
Coulomb explosion imaging (CEI) has become prevailing to determine the molecular structure by measuring the momentum vectors of fragmented ions following multiple ionization processes in the pump-probe experiments~\cite{Cooper21:JPCA125,Gagnon08:JPB41}, and one can map the measured fragment momentum data to the instantaneous geometry of the molecule at the time of the ionization. The CEI method is widely accepted because it allows convenient interpretation of the results, e.g. Coulomb explosion of methanol reveals its complex mechanisms for different production breakup channels including transient proton migration and long-range "inverse harpooning"~\cite{Luzon19:JPCL10},
the conformational isomers of 1,2-dibromoethane (C$_2$H$_4$Br$_2$) can be distinguished by exploiting the three-body breakup channel $\mathrm{C}_2\mathrm{H}_4^++\mathrm{Br}^++\mathrm{Br}^+$ containing both sequential and concerted breakup processes~\cite{Pathak20:JPCL11}. 
However, the accurate reconstruction of molecular structure could hardly go beyond the measurement of bond length of diatomic molecules, because the complex atomic motion during the Coulomb explosion leads to the deviation of measured bond angle. And this in most cases hinders direct reconstruction of molecular structure, unless a full time reversed propagation of molecular fragment trajectories is made possible. The required accuracy of momenta measurement is not within reach of current experimental techniques~\cite{ZL17:NC8}.

{Using tabletop lasers, x-ray diffraction has found many applications in determination of crystalline structural dynamics, such as the spatially periodic femtosecond excitation of GaAs/AlGaAs superlattice~\cite{Bargheer04:Science306} and acoustic
phonon dynamics in a Ge/Si layered structure~\cite{Cavalleri00:PRL85}. }
The development of ultrabright electron sources and the advent of x-ray free electron lasers (XFEL) broaden the application scope for imaging of molecules{~\cite{Ware19:PRA100,Minitti15:PRL114,Budarz16:JPB49,Yong18:JPCL9}}. Compared to Coulomb explosion imaging, ultrafast coherent diffraction imaging provides an opportunity to directly image the molecular configuration and structural dynamics~\cite{ACSPhot20:ZL7,Ischenko17:CR11066}.  
High intensity and short pulse length of 
XFEL can {enable diffraction sample to be isolated molecule of gas phase} and reconstruct the image of non-crystalline objects~\cite{Chapman09:Natmat8} by detecting continuous diffraction intensity rather than Bragg spots.
{For example, one- and two-photon dissociation pathways of idione molecule are identified from the frequency-resolved x-ray scattering, which reflects two different rates of increase of internuclear distance~\cite{Ware19:PRA100}. }

Ultrafast electron diffraction (UED) exhibits high spatial resolution, but suffered from space
charge broadening which limits the temporal resolution~{\cite{Martin16:JPB49}}, the issue is resolved by using rf pulse compression and relativistic effect of electrons of MeV energy, and the UED can currently have the time resolution of $\sim$50fs~\cite{ACSPhot20:ZL7,YangJ20:Science368}. 
Compared to x-ray scattering, electron scattering is more sensitive to the properties of electronic states such as electronic correlation, and goes beyond the independent atom model commonly used for traditional diffraction experiments, which neglects the redistribution of valence electrons during chemical-bond formation~\cite{Santra09:JPB42}. For example, the S$_1(n\pi^*)$ state population evolution of pyridine is reflected by inelastic signal related to the two-electron correlation effect due to Coulomb repulsion between electrons~\cite{YangJ20:Science368}.
Starting from the probability densities of {molecular rotational wavepacket determined by ultrafast diffraction~\cite{Xiong20:PRR2}}, the density matrix of the quantum wave packet can be obtained by quantum tomography~\cite{MZ21:NC12}.

In this Perspective, we emphasize the ultrafast imaging methods illustrated in Fig.~\ref{fig:roadmap}, which make use of table-top laser, XFEL and electron sources.

\begin{figure}
    \centering
    \includegraphics[width=12cm]{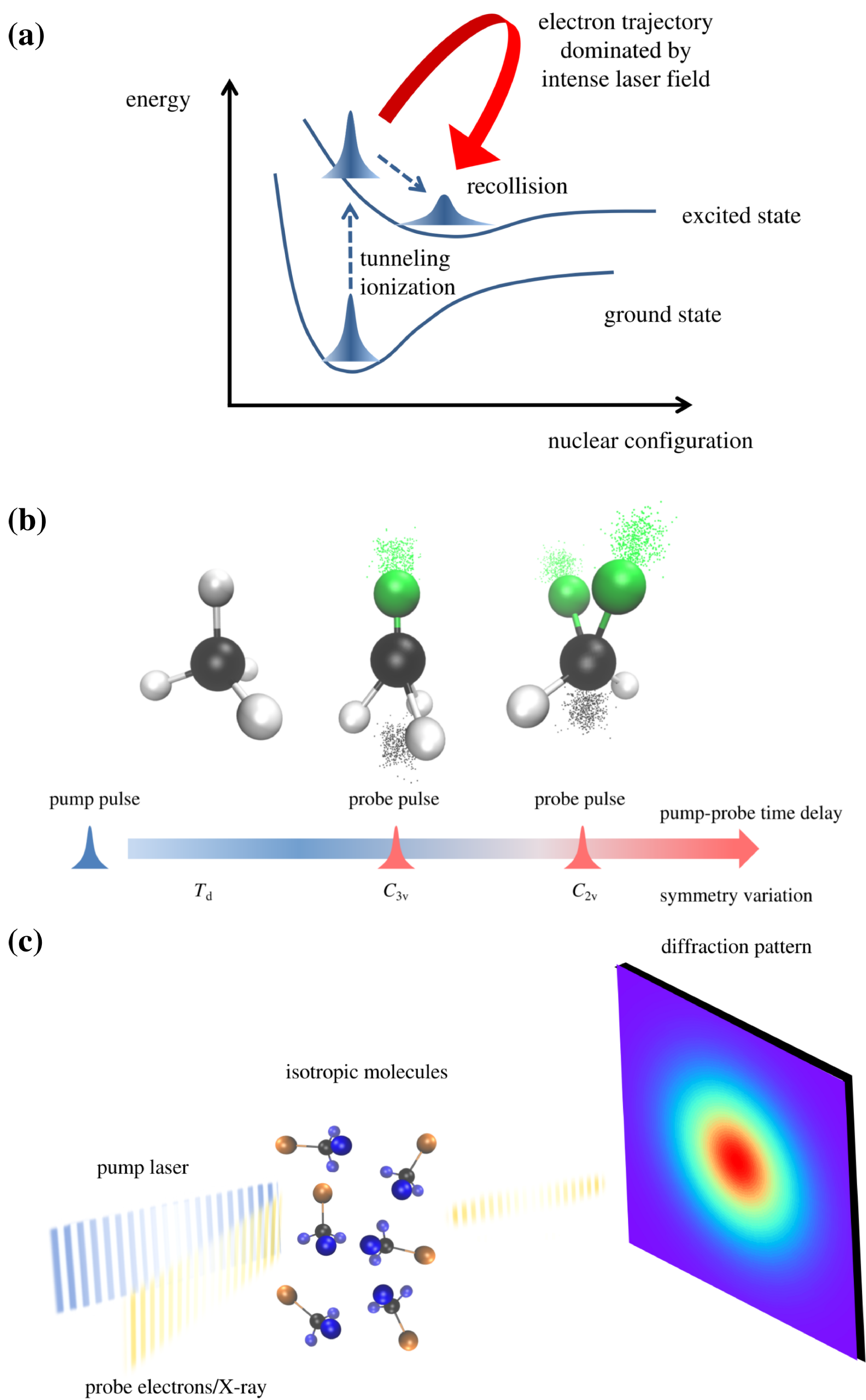}
    \caption{Schematic figure of strong laser field tunneling ionization, recollision-based imaging, Coulomb explosion imaging and diffraction imaging. 
    (a) High harmonic generation imaging and photoelectron imaging are based on the recollision between the tunnel-ionized electron and parent ion. (b) Symmetry variation of methane cation is reflected by the changing of Coulomb explosion dissociation channels detected by different pump-probe time delay. (c) Gas-phase electron/x-ray diffraction imaging of molecules encodes structural and electronic information.}
    \label{fig:sch}
\end{figure}

\textbf{Ultrafast Imaging by Laser-induced Ionization.} In the prototypical strong-field interaction, the valence electron can be released through tunnel ionization from the molecule. The photoelectron momentum distributions (PMDs) encode structural
information of its origin. The measured far-field PMDs can be understood as a diffraction image of the source. Thus, in principle, it should be possible to retrieve structural information by analyzing the interference pattern. However, the source is not identical to the atomic or molecular orbital from which the electron is removed but rather the corresponding Dyson orbital. 
For the molecular species, the molecular frame photoelectron angular distributions can be measured. By measuring the PMDs of the aligned molecules N$_2$, O$_2$~\cite{Meckel08:Science320,Meckel14:Natphy10,MML16:PRL116}, the molecular structures have been imaged. Recently, the strong-field direct ionization of the complex molecules has been demonstrated ~\cite{Wiese21:PRR3}.
The photoelectron spectroscopy of CS$_2$ reflects its nonadiabatic photodissociation processes involving vibrational quantum beat and different dissociation channels through singlet and triplet intermediate metastable states~\cite{Karashima21:JPCL12}.
In the future, when the complex molecules are spatially confined, it would be possible to extract the structure for the complex molecules.
Apart from detecting nuclear dynamics, intriguing approaches are recently developed to investigate ultrafast electron dynamics in molecules, such as the attosecond angular streaking (attoclock) with circular femtosecond laser fields, in which the photoelectrons can be streaked into different direction depending on the ionization instant in the molecular frame. 
The geometry of two-color co-rotating circularly polarized fields has been used as the novel attoclock scheme for atoms~\cite{Manhan:PRL18,Peipei:PRL20}. 
Transferring this technique from atoms to polyatomic molecules could provide a potential approach to study the ultrafast laser-molecular interaction and electron dynamics of molecules.

\textbf{Imaging Methods using Laser-induced Rescattering.} As shown in Fig.~\ref{fig:sch}(a), besides direct electron tunneling ionization, the strong laser field can also accelerate the free electron in vacuum and eventually drives it back to the parent ion, predominantly resulting in rescattering or radiative recombination. The radiative recombination results in the emission of high-energy photons through high-harmonic generation. The three steps of HHG in semiclassical model~\cite{Corkum93:PRL71} can be correlated with a pump-probe process: the tunneling ionization plays the role as the pump that launches nuclear dynamics, and recollision of electron and parent ion is the probe step for nuclear motion occurring during the time delay~\cite{baker06:424}. When the fraction of ionized electron wavepacket interferes with the orbital from which the electron was extracted, the electron density distribution oscillates as the wave packet propagates and leads to radiative emission of harmonics~\cite{Ville04:Nature432}. By aligning the molecules, the spectrum of high harmonic generation can be measured, it is a powerful tool to investigate the electronic structure with attosecond temporal resolution. Alternatively, the rescattered portion of the electron wavepacket is exploited in laser-induced electron diffraction experiments as a coherent diffraction pattern of the molecular target. 
{For example, LIED is used to retrieve multiple bond lengths of polyatomic molecule acetylene (C$_2$H$_2$)~\cite{wolter16:Science308,Pullen15:7262}. The cleavage of C-H bond and full structure of di-ionized acetylene during its dissociation dynamics can be visualized by LIED for both parallel and perpendicular oriented case~\cite{wolter16:Science308}}.
C. I. Blaga $et$ $al.$ use gas-phase LIED to map the structural responses of oxygen and nitrogen molecules to ionization, reaching the spatio-temporal resolution of 0.1-\AA~change in the oxygen bond length occurring in the time scale of $\sim$5 fs~\cite{Blaga12:194}. Several challenges arise for experimental techniques including achieving high recollision energies and sufficient momentum transfer for the ionized electron~\cite{Pullen15:7262}. 
In contrast with the conventional electron diffraction, the influence of strong laser field needs to be removed for extraction of field-free electron–ion cross sections based on the quantitative rescattering theory~\cite{Lin10:JPB43}. 
In the future, the application of laser-induced rescattering effect can potentially provide time-dependent images of the molecule at sub-femtosecond and few-picometer resolution for the complicated molecules using the mid-infrared or infrared lasers and the higher pondermotive potential.
 
\textbf{Coulomb Explosion Imaging.} The Coulomb explosion imaging method provides instantaneous 3D imaging of nuclear positions~\cite{Vager01:203}, such as the structure of hydrogen dimers~\cite{Khan20:JPCL11}, laser-induced isomerization in acetonitrile (CH$_3$CN)~\cite{McDonnell20:JPCL11} and photodissociation dynamics of chlorocarbonylsulfenyl chloride~\cite{Cooper21:JPCA125} and CH$_2$BrI~\cite{Burt17:PRA96}.
In CEI experiments, the bonding electrons of molecules are stripped away by ultrashort lasers in a time scale shorter than nuclear motion, leaving positive ions in the positions of molecular nuclei. High intensity lasers is feasible to stripping many electrons, and typically, at least several binding electrons are scattered away. For visible laser, sufficiently high intensity ($\sim 10^{14}$ W/cm$^2$) is needed to successively remove the electrons from molecule and generate the interatomic Coulomb repulsion~\cite{Corrales12:JPCA116,Wang08:JPCA112}. In contrast, core ionization by x-ray will lead to cascading Auger decay and create highly charged molecules, which fall apart under Coulomb repulsion~\cite{Liek6:NC15}. As the molecular potential suddenly changes to Coulomb repulsion dominated potential of the ion's degree of freedom, the molecule dissociates and the CEI detector collects velocities by simultaneous determination of molecular fragments from single molecule, and the 3D images of nuclear positions can be resolved by analyzing the momentum distribution of the fragments. The fragment trajectories for predicting detected velocity from initial conformations can be approximated using classical simulations with sufficient accuracy when at least one of the fragments charge is two or more. The density function of conformations including correlated variations within the molecular structure can be obtained by collecting structures of a molecular ensemble. The scheme of obtaining molecular structure from CEI is relatively direct and model independent, and classical trajectories approximation is usually sufficient especially for highly charged fragments. But the sudden approximation of electron stripping may not always be not fully satisfied, leading to deviation from this simplified interpretation~\cite{Vager01:203}. 

As shown in Fig.~\ref{fig:sch}(b), by extracting the atomic motion with Coulomb explosion momentum mapping, the variation of molecular symmetry variation processes can be revealed. A prominent example is the spontaneous symmetry breaking, which is a fundamental phenomenon occurred in numerous fields of physics, such as the Higgs mechanism~\cite{Higgs64:PRL13}, and Jahn-Teller effect in molecular systems. The Jahn-Teller (JT) effect, arising from nonadiabatic coupling between electrons and nuclei, leads to distortion of nuclear configuration away from high symmetric points accompanied by removal of electronic state degeneracy~\cite{Bersuker06:JTE}. For a typical linear vibronic coupling Hamiltonian of two state, the diabatic potential forming JT type conical intersection can be written as~\cite{Arn18:PRL120}
\begin{equation}
    V=\begin{pmatrix}
    V_1(x,y) & \lambda y \\ \lambda y & V_2(x,y)
    \end{pmatrix}
    \,,
\end{equation}
where the tuning mode $x$ and coupling mode $y$ are two nuclear coordinates, and $V_1$ and $V_2$ are two diabatic state potential energy surface. Experimentally, the JT deformation of methane cation could be indirectly imaged by Coulomb explosion imaging in $\sim$10 fs time scale using tabletop laser, which is able to create ions up to triple charges by strong field ionization~\cite{Min21:NC12}. The strong pump pulse ionizes neutral methane and populates the CH$_4^+$ ground state $^2F_2$. After JT distortion from $T_{\text{d}}$ to $C_{\text{3v}}$ geometry, $f_2$ and $e$ vibrational modes lead to further distortion to $C_{\text{2v}}$ symmetry, where the electronic degeneracy is completely lifted~\cite{Jacovella18:JMS343}. A $\sim$20 fs time delay is observed between the two-body and three-body Coulomb explosion channels, which corresponds to the symmetry evolution from $C_{\text{3v}}$ to $C_{\text{2v}}$, and is driven by the superposition of multiple vibrational modes~\cite{Min21:NC12}.
However, the limitation of low cross sections in creating highly charged molecular ions through strong field valence ionization by tabletop lasers with wavelength in the IR to visible regime hinders the accuracy of structural analysis in CEI. Ideally, we expect to have an $n$-ion for an $n$-atomic molecule, in this manner, we can retrieve maximal information about the atomic positions in the molecule at the instant of Coulomb explosion.
\begin{figure}
    \centering
    \includegraphics[width=16cm]{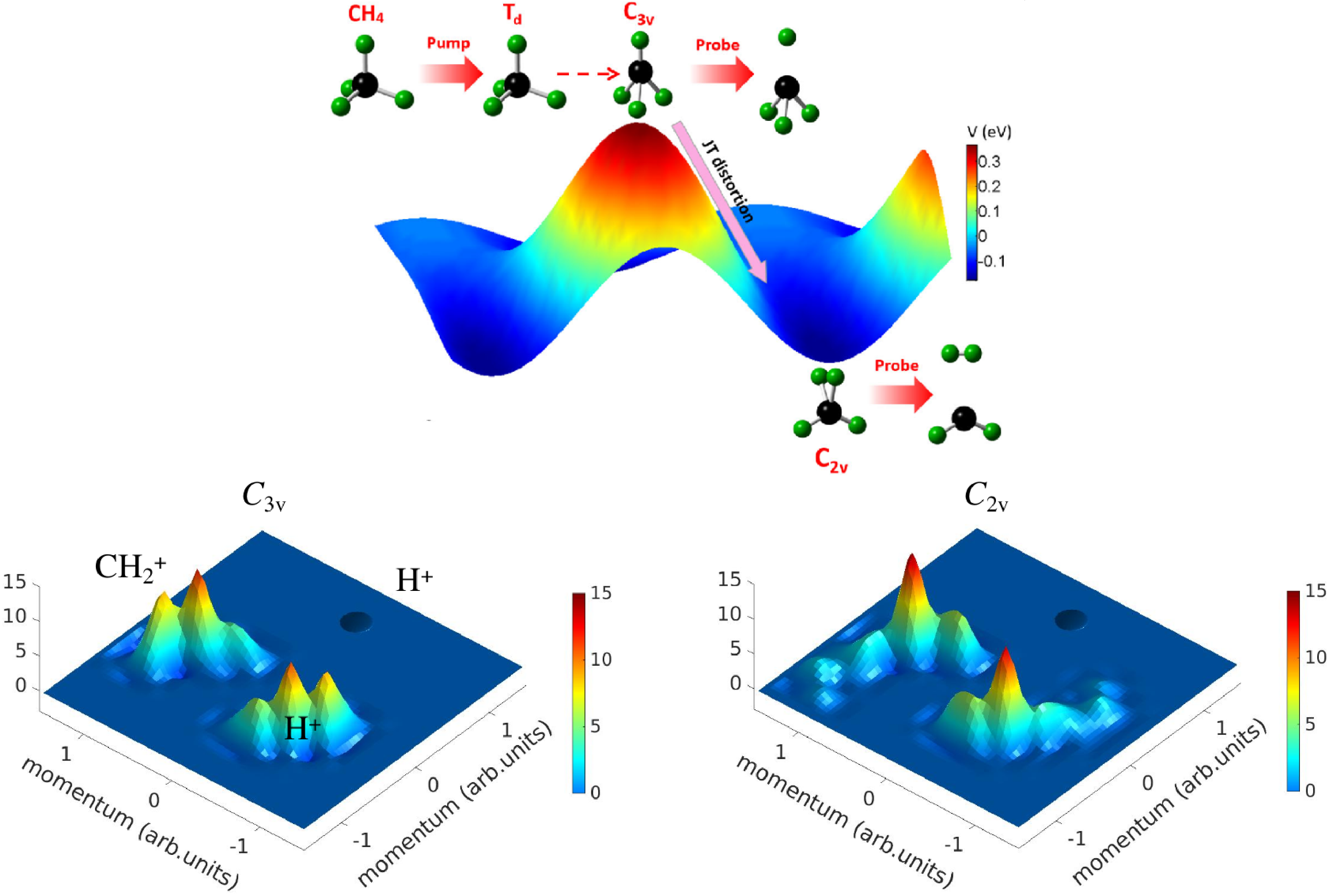}
    \caption{Potential energy surface (PES) for $\rm{CH}_4^{+}$ cation and Newton map of three-body Coulomb explosion momentum mapping at 9 fs and 29 fs pump-probe time delay. The $\rm{CH}_4^{+}$ cation in a $C_{\text{3v}}$ geometry (at the peak of PES) undergoes JT distortion to a $C_{\text{2v}}$ geometry (at the bottom of PES). The spot-like structures correspond to different molecular symmetries of CH$_4^+$. The momentum mapping of reference configuration $C_{\text{3v}}$, $C_{\text{2v}}$ and $D_{\text{2d}}$ is based on the simulated result of Coulomb repulsion, which takes into account the atomic motion during CEI process.  }
    \label{fig:ch4}
\end{figure}

One of the solutions is to use ultrafast x-ray as probe to induce Coulomb explosion, this has been realized by free electron lasers, which can easily create highly charged molecular states via ionization cascades~\cite{ZL17:NC8}. As an example, the proton migration process for acetylene dication can be revealed by x-ray CEI, which is related to the nature of chemical bonding forming and breakup. {The molecule can be exploded into more fragments for higher charged state and more detailed information of molecular structure can be detected. For the case of C$_2$D$_2$, the Coulomb explosion of $4+$ ions provides complete information of each atom. By contrast, the three-body Coulomb Explosion channel shown in Fig.~\ref{fig:ch4} induced by intense laser can not provide the structural information inside the CH$_2^+$ fragment.} X-ray probe creates C$_2$D$_2^{4+}$ ion from C$_2$D$_2^{2+}$ dication for Coulomb explosion, and the CEI exhibits increase of average C-C-D angle with temporal evolution, which resolves the bending motion due to softening of the $\sigma$ bonding of C-C-D atoms within 100 fs in $^1\Sigma^+_g$ state~\cite{Liek6:NC15}. However, even in case with Coulomb explosion of $4+$ ions for the four-atomic molecule, the disadvantage of CEI as an indirect imaging method prevents it to resolve the molecular dynamics by experiment alone and one has to incorporate input from theoretical simulations.
In the case of CEI of C$_2$D$_2^{2+}$ dynamics, the ab initio calculation predicts much longer timescale due to an isomerization barrier of $\sim$2eV~\cite{ZL17:NC8}. Actually, the signal can be reproduced by the relative rotation of Coulomb exploded fragments, but large magnitude proton motion without leading to isomerization does exist on sub-100 fs timescale because of $\sim 2.4$ eV relaxation energy released from the weakening of C-C bond in high-lying $1\pi_u^{-1}3\sigma_g^{-1}$ states~\cite{ZL17:NC8}.   The enhancement of momentum measurement accuracy of provides the possibility to resolve the rotational motion of Coulomb explosion fragment and to directly reconstruct molecular structure by time reversed propagation of the Newtonian equation of motion. However, the required resolution can be exceptionally high. For the dynamics of C$_2$D$_2^{2+}$ dication, the needed resolution can be estimated by considering a bending mode of vibrational period $T\sim 100$ fs and of bond length $r\sim 1$ \AA. Due angular momentum conservation during the Coulomb explosion process, the tangential momentum $p_{\bot}$ perpendicular to the bond explosion direction at the detector position $R\sim 0.1$ m is given by  ${2\pi}mr^2/T=Rp_{\bot}$, which gives a required accuracy $p_{\bot}\sim 10^{-8}$ a.u., far above the current experimental resolution of momentum measurements. 

\textbf{Ultrafast X-ray Diffraction Imaging.} Ultrafast coherent diffraction imaging provides a direct access to resolve the dynamical molecular structure, and thus offers an alternative route that is complementary to the Coulomb explosion imaging.  {Beyond static x-ray diffraction measurement such as gas-phase molecule like strongly aligned 2,5-diiodothiophene~\cite{Kierspel15:JPB48} molecules, atomic motions and molecular dynamics can be revealed by time-resolved x-ray diffraction.}
As shown in Fig.~\ref{fig:sch}(c), the electronic information and the atomic positions in the molecule are encoded in the scattered wave of coherent incident x-ray~\cite{Bartell64:JACS86}, and can be determined by the measurement of diffraction intensity. The lost phase information of the electron form factor can be recovered by phase retrieval algorithms~\cite{Marchesini07:RSI78}.
{According to the time-resolved nonresonant x-ray scattering theory~\cite{Dixit14:PRA89,Dixit12:PNAS109,PhysRevB.98.224302,Dixit17:PRA96}, it was shown that ultrafast x-ray diffraction of electron wavepacket in isolated atoms and molecules is in fact related to the electron density correlation function, which does not intuitively correspond to the Fourier transform of instantaneous electron density because inelastic x-ray scattering can be play important role when the indication of elastic scattering by Bragg spots is not available ~\cite{Dixit12:PNAS109}, it makes the ultrafast diffraction of isolated atoms and molecules unique from traditional physical picture. Numerous theoretical studies have been focused on reconstructing nuclear and electronic dynamics from time-resolved x-ray diffraction, such as lattice dynamics of germanium~\cite{Rana21:NPJ7}, rotational dynamics in linear molecule NaF~\cite{Simmer20;PRA102}, valence electronic dynamics in oxazole~\cite{YongH21:JPCL12} and during a pericyclic reaction of semibullvalene~\cite{Bredtmann14:NC5} entangled valence electron-hole dynamics~\cite{Healion12:JPCL3}, transient electronic coherences of H$_2$~\cite{Simmer19:PRL122}, charge migration in chiral molecule epoxypropane~\cite{Giri21:PRA104} and electron current flux density of benzene~\cite{Hermann20:PRL124}.}

For crystalline samples, the diffraction signal at discrete Bragg peaks comes from interference enhancement of scattered waves from different lattice sites $\vec{R}_\alpha$. The total charge density in Fourier space $\hat{\sigma}=\sum_{\alpha}\hat{\sigma}_{\alpha}=\sum_{\alpha}\hat{\sigma}e^{-i\vec{s}\cdot\vec{R}_\alpha}$ is the sum of individual charge density, where $\vec{s}$ is the momentum transfer between incident and diffracted x-ray photon. The total diffraction intensity (apart from a constant coefficient) for $N$ lattice sites is~\cite{Kochise17:069301}
\begin{equation}
    I=\mathrm{Tr}^{(N)}(\hat{\rho}^{(N)}\hat{\sigma}^{\dagger}\hat{\sigma})=\sum_{\alpha=\beta}\mathrm{Tr}^{(1)}(\hat{\rho}^{(1)}_{\alpha}\hat{\sigma}_{\alpha}^{\dagger}\hat{\sigma}_{\alpha})+\sum_{\alpha\neq\beta}\mathrm{Tr}^{(2)}(\hat{\rho}^{(2)}_{\alpha\beta}\hat{\sigma}_{\alpha}^{\dagger}\hat{\sigma}_{\beta}),
\end{equation}
where the atoms or molecules from different lattice sites are uncorrelated, so the second order density matrix of two sites is direct product of the density matrix for each site $\hat{\rho}^{(2)}_{\alpha\beta}=\hat{\rho}^{(1)}_\alpha\otimes\hat{\rho}^{(1)}_\beta$. Suppose the excitation ratio from ground state $\ket{\psi_g}$ to excited state $\ket{\psi_e}$ is $\eta$,
\begin{eqnarray}
    \hat{\rho}^{(1)}_\alpha=\eta\outerproduct{\psi_e}{\psi_e}+(1-\eta)\outerproduct{\psi_g}{\psi_g}\,.
\end{eqnarray}
The first term of diffraction intensity is the unimolecular contribution,
\begin{align}\nonumber
    I_1&=\sum_{\alpha=\beta}\mathrm{Tr}^{(1)}(\hat{\rho}^{(1)}_{\alpha}\sigma_{\alpha}^{\dagger}\sigma_{\alpha})\\\nonumber
    &=\sum_{\alpha}\left[\eta\mathrm{Tr}^{(1)}(\outerproduct{\psi_e}{\psi_e}\hat{\sigma}^{\dagger}\hat{\sigma})+(1-\eta)\mathrm{Tr}^{(1)}(\outerproduct{\psi_g}{\psi_g}\hat{\sigma}^{\dagger}\hat{\sigma})\right]\\
    &=N\left[\eta\braopket{\psi_e}{\hat{\sigma}^{\dagger}\hat{\sigma}}{\psi_e}+(1-\eta)\braopket{\psi_g}{\hat{\sigma}^{\dagger}\hat{\sigma}}{\psi_g}\right]\,,
\end{align}
for the electron form factor $f_{g/e}=\braopket{\psi_{g/e}}{\hat{\sigma}}{\psi_{g/e}}$ of the ground and excited state. Ignoring transition density terms $\braopket{\psi_g}{\hat{\sigma}}{\psi_e}$, we have the elastic unimolecular signal
\begin{align}
    I_1=N\left[(1-\eta)|f_g|^2+\eta|f_e|^2\right]\,.
\end{align}
The biomolecular contribution is
\begin{align}
I_2&=\sum_{\alpha\neq\beta}\mathrm{Tr}^{(2)}(\hat{\rho}^{(1)}_{\alpha}\hat{\sigma}_{\alpha}^{\dagger}\otimes\hat{\rho}^{(1)}_{\beta}\hat{\sigma}_{\beta})\nonumber\\
&=\sum_{\alpha\neq\beta}e^{i\vec{s}\cdot(\vec{R}_{\alpha}-\vec{R}_{\beta})}\mathrm{Tr}^{(1)}(\hat{\rho}^{(1)}\hat{\sigma}^\dagger)\mathrm{Tr}^{(1)}(\hat{\rho}^{(1)}\hat{\sigma})\nonumber\\
&=\sum_{\alpha\neq\beta}e^{i\vec{s}\cdot(\vec{R}_{\alpha}-\vec{R}_{\beta})}\left|(1-\eta)f_g+\eta f_e\right|^2\,.
\end{align}
The diffraction signal $I_2$ scales as $N^2$, which is overwhelmingly predominant in the crystal diffraction, because $I_1$ is proportional to $N$. The interference of ground state and excited state form factors appeared in $I_2$ amplifies the change of diffraction signal, especially for weak excitation $\eta\ll 1$. 
For example, the photoexcitation of polycrystalline ammonium sulfate [(NH$_4$)$_2$SO$_4$] induces electron transfer followed by a geometry stabilization process. By measuring the change of x-ray diffraction intensity of Bragg peaks $\Delta I_{hkl}$, the electron flow pathways and proton translocation processes can be determined, which are reflected by the modification of structural factors~\cite{Woerner10:064509}. 
{The structural relaxation of N,N-dimethylaminobenzonitrile following photoexcitation is observed by time-resolved x-ray powder diffraction of with a relaxation lifetime of 520 ps~\cite{Tech01:PRL86}. As another example, ultrafast x-ray diffraction allows for monitoring the structural switching process of the photoreaction of Photochromic $\alpha$-Styrylpyrylium Trifluoromethanesulfonate in crystal thin film~\cite{Hall09:JACS131}. 
In protein crystallography, synchrotron radiation sources is limited by the requirement for macroscopic crystals, which can be difficult to obtain. However, XFEL is capable of delivering extremely intense femtosecond pulses to either macroscopic or microscopic crystals for the structure determination of undamaged macromolecules~\cite{Schlichting15:IUCrJ2}. 
This XFEL-based time-resolved crystallography is applied to understand the allostery and dynamics of fluoroacetate dehalogenase over the course of four full catalytic turnovers. It collects snapshots along reaction coordinates in the catalystic cycle and demonstrates the allostery happens without conformational change associated with substrate binding and catalysis~\cite{Mehrabi19:Science365}. 
As a variant of time-resolved surface X-ray diffraction (TRXD), time-resolved surface X-ray diffraction~\cite{Robinson92:RPP55} at grazing incidence develops a lot with the advent of XFEL. It has been experimentally demonstrated that TRXD is capable to selectively map the motion of valence electrons in an adsorbate-surface system~\cite{Toepfer2021:JPCC125}. 
Using a smooth limited-$Q$ Fourier filter, which can be understood as the convolution of the detected signal with a finite box function, and a Gaussian-smoothed cutoff apodization function, the valence electrons' motion of $\mathrm{Cu}_2$/$\mathrm{MgO}(100)$ are reconstructed.}

For diffraction from gas phase molecules, the average of bimolecular diffraction $I_2$ vanishes due to randomly distributed molecules and the only detected signal is $I_1$.  
X-ray free electron laser can satisfy the high intensity requirement and {generalize diffraction sample to noncrystalline cases and even single molecule limit~\cite{Starodub12:NC3,Kahra12:NP8}}. Ultrafast laser pulses shorter than the time scale of the radiation damage enable the room temperature "diffraction before destruction" technique for {isolated molecules of gas phase} diffraction imaging~\cite{Barty13:ARPC64} and provide high temporal resolution to image dynamical processes of molecules with femtosecond resolution. 
For strongly aligned 2,5-diiodo-benzonitrile molecule, the coherent x-ray diffraction pattern arising from the interference of two iodine atoms at both ends separating by 700 pm is observed using 2 keV (620 pm) XFEL~\cite{Kupper14:PRL112}. Although the first interference maximum is not recorded due to relative long wavelength, the distance of two iodine atoms can be derived from measured diffraction intensity by $\chi^2$ analysis with simulation results. A more precise result showing the first three maxima of iodine-iodine interference is obtained by increasing the photon energy to 9.5 keV~\cite{Kierspel20:084307}.
{ Ultrafast x-ray diffraction also provides wide application to molecular structural dynamics.
The initially excited state of N-methyl morpholine can be identified as 3$p_z$ Rydberg state by determining the orientation of transition dipole moment from anisotropy of x-ray diffraction signals~\cite{Yong18:JPCL9}. Furthermore, its temporal evolution measured from x-ray scattering reflects vibrational coherent motion which persists for multiple vibrational periods~\cite{Stankus19:NC11}.
During the ring-opening reaction of 1,3-cyclohexadiene (CHD), as a typical example of electrocyclic reactions, the evolving dynamics such as the change of distances between non-neighboring carbon atoms can be reflected by time-dependent x-ray diffraction intensity with 80 fs time scale~\cite{Minitti15:PRL114,Budarz16:JPB49}.}

\begin{figure}
    \centering
    \includegraphics[width=15cm]{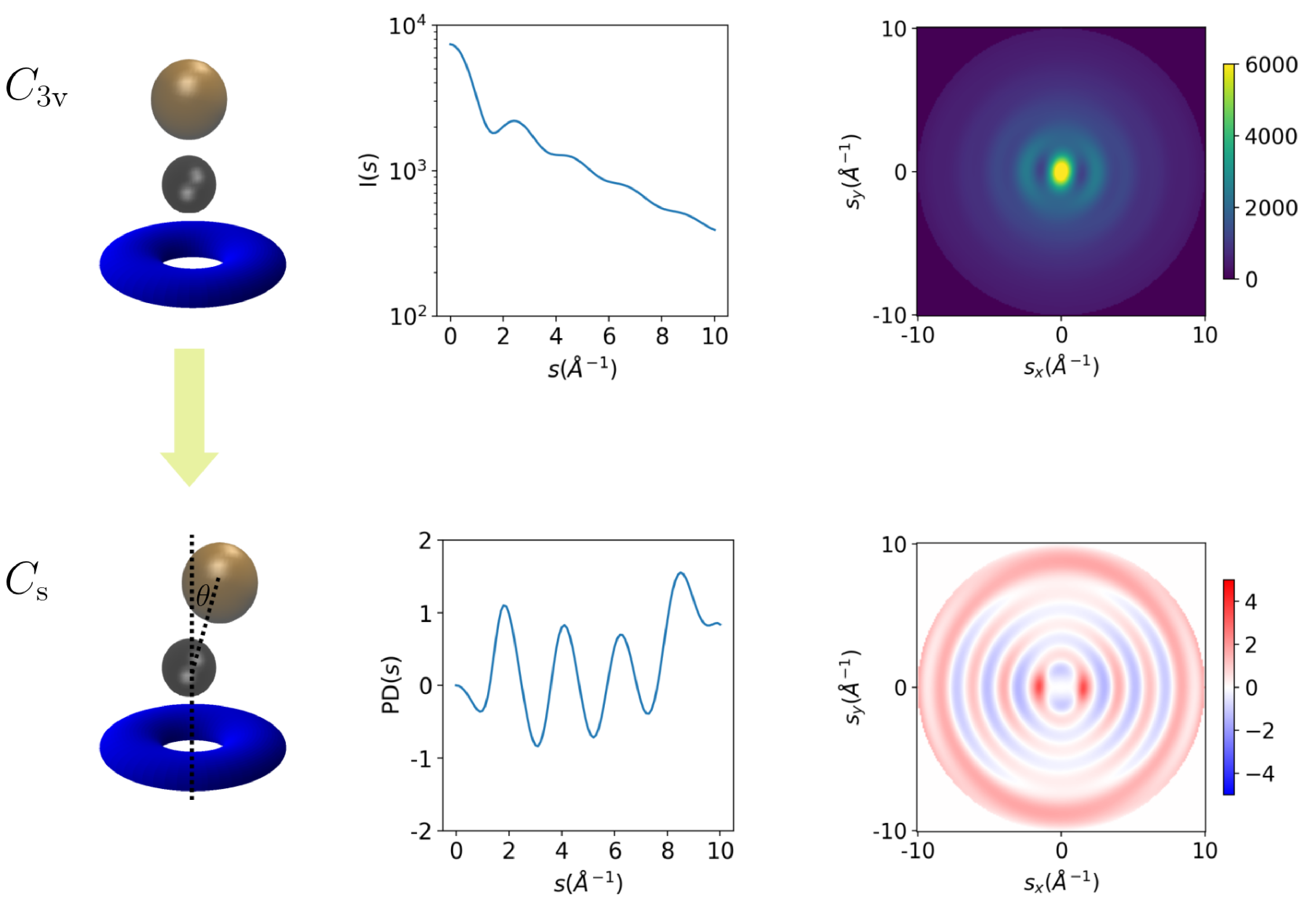}
    \caption{Rotationally averaged CF$_3$I diffraction pattern of $^3E$ states. Brown and black spheres are I and C atom, respectively, and the blue torus represents three F atoms. 1D and 2D x-ray diffraction pattern of Franck-Condon geometry is shown in the upper panel (arbitrary unit), and percentage difference (PD) diffraction pattern of JT distortion geometry ($\theta=4.5^\circ$) is shown in the lower panel. }
    \label{fig:Difpat_cf3i}
\end{figure}

However, the contemporary spatio-temporal resolution of ultrafast diffraction is, in certain cases, still insufficient to detect molecular dynamics of small amplitude motion, such as the Jahn-Teller structural distortion of photoexcited CF$_3$I molecules. We simulated the diffraction pattern of CF$_3$I before and after the JT distortion on the CASSCF(6,5)/DZVP level using TeraChem program~\cite{Ufim09:JCTC5}. Because of the JT effect (see Fig.~\ref{fig:Difpat_cf3i})~\cite{Ama91:JCP94}, the bending of C-I bond lifts the degeneracy of $^3E$ states from $C_{\text{3v}}$ to $C_{\text{s}}$ geometry, and the distortion angle for C-I bond bending $\theta=4.5^\circ$ corresponding to the minimum of $A'$ state potential energy curve. As the C$_{\text{3v}}$ symmetry breaks, the iodine atom fragments on a conical surface around the molecular axis, and generates an off-axis recoil to the CF$_3$ group. The diffraction pattern and the percentage difference (PD) of signal between Franck-Condon (FC) geometry and JT geometry defined as~\cite{YangJ20:Science368}
\begin{equation}
    \text{PD}(\textbf{s})=\frac{I_{\text{JT}}(\textbf{s})-I_{\text{FC}}(\textbf{s})}{I_{\text{FC}}(\textbf{s})}\times 100
    \,.
\end{equation}
are shown in Fig.~\ref{fig:Difpat_cf3i}. 
The percentage difference signal of diffraction pattern reflects the change of electronic density distribution and the molecular structure variation.

{However, the signal does not show distinctive features that can be identified as JT distortion. In the ultrafast electron diffraction experiment of CF$_3$I~\cite{YangJ18:Scicen361}, the JT distortion has not been measured from experimental data. For example, during the dissociation processes, the distances between different fragments change significantly, which can be directly reflected by the slope of frequency-resolved x-ray scattering signal indicating the dissociation velocity~\cite{Ware19:PRA100}, or the change of pair distribution functions reflecting the distance of atom pairs obtained from ultrafast electron diffraction~\cite{YangJ18:Scicen361}. The small amplitude JT distortion, although manifests itself by around 1\% amplitude signal as shown in Fig.~\ref{fig:Difpat_cf3i}, can not be directly imaged if not comparing with calculation results.}

This pushes us to envisage new theory for ultrafast diffraction beyond classical ball-and-stick model of molecules, which can assist us to tackle open problems such as resolving small amplitude JT distortion. A savvy way to reveal Jahn-Teller effect in photoexcited CF$_3$I is to use the imprints of symmetry breaking left in the quantum state of molecules. During the C-I bond breakup process, the off-axis recoil momentum of the I atom acting on the CF$_3$ fragment can obviously lead to its rotational excitation~\cite{Kos00:1009}. Reconstructing the quantum mechanical density matrix of the excited rotational states of CF$_3$ is equivalent to resolving the JT effect. 
Because the ultrafast coherent diffraction encodes the information of time-dependent probability density of the molecular quantum system, it naturally provides the possibility for tomographic reconstruction of the quantum states~\cite{Ischk13:PRI13,Ischk99:SPIE16}. Restricted by the dimension problem~\cite{Mouritzen05:PRA,Mouritzen06:JCP244311}, traditional quantum tomography (QT) method~\cite{Leonhardt96:PRL1985,Dunn95:PRL884} can only be applied to systems of a single degree of freedom, which categorically excludes the rotation problem. However, the underlying nature of QT is to retrieve the phase information lost in measurement~\cite{ACSPhot20:ZL7}. Analogous to crystallographic phase retrieval, the dimension problem can be solved by iterative projection algorithm using additional physical constraints of density matrix~\cite{MZ21:NC12}. The generalized QT algorithm resolves the dimension problem and can be applied to make quantum version of "molecular movie" for rotational and vibrational motions.

\textbf{Ultrafast Electron Diffraction Imaging.} Compared with x-ray diffraction, electron diffraction exhibits higher spatial resolution due to its shorter de Broglie wavelengths and larger scattering cross sections, which is especially important for the diffraction experiments~\cite{wolter16:Science308}. {In liquid-phase and gas-phase UED experiment, conventionial keV-UED can resolve photoproducts and long life intermediates but is incapable of resolving atomic motions due to temporal resolution~\cite{Cao99:PNAS96,Hyotcherl01:Science291}. In recent years, with MeV electrons,} gas phase ultrafast electron diffraction can provide sub-\AA~spatial resolution and sub-picosecond temporal resolution using compact setups~\cite{Martin16:JPB49}. 
{Electron diffraction of solid state samples are ideal to understand collective effects in condensed matter, such as insulator-to-metal phase transition of organic salt (EDO-TTF)$_2$PF$_6$~\cite{GaoM13:Nat496} and solid-to-liquid phase transition caused by disordering in bismuth~\cite{Sciaini09:Nature458}. However, most molecular crystals are insulators with low melting points, and low repetition rates must be used to avoid cumulative heating effects~\cite{Jean13:JPCB117}. Gas-phase electron diffraction images prototypical reactions in chemistry and can be directly linked to quantum chemical calculations for isolated molecules~\cite{YangJ16:NC7}. But molecular alignment is often required because the random
orientation of molecules often limits the ability to describe distributions of internuclear distances in the gas phase~\cite{Ischenko17:CR11066}.}

The spatial charge broadening due to Coulomb repulsion of electrons and group velocity mismatch between electron and laser pulses decreases the temporal resolution~\cite{YangJ16:PRL117}, which can be remedied by relativistic electron acceleration or electron bunch compression~\cite{Yingp15:JPCL6,Gliserin15:NC6}.
Because the longitudinal space-charge pulse elongation is proportional to $1/\beta^2\gamma^5$, where $\beta=v/c, \gamma=(1-\beta^2)^{-1/2}$ and $v,c$ are the speed of electrons and vacuum light speed, respectively, the spatial charge effect can be greatly diminished for relativistic electrons~\cite{Martin16:JPB49,YangJ16:NC7}. {MeV electron diffraction based on an $S$-band photocathode rf gun pulse compression can achieve 100 fs temporal resolution of polycrystalline diffraction aluminium foil~\cite{Li09:Rsi80} and 65 fs temporal resolution for gas-phase diffraction of CF$_3$I~\cite{Shen19:Struct6}, which significantly improved the application of gas-phase and crystalline MeV ultrafast electron diffraction. 
The prospect of reaching 10 fs temporal resolution by relativistic electron sources~\cite{Manz15:FD177,Shin21:LPR15} has great potential for the exciting development of mapping atomic motions and molecular dynamics. }

Another advantage of electron diffraction is its sensitivity in detecting electronic correlation, which is incorporated in the inelastic part. This can be understood by the difference in the scattering operators of x-ray diffraction $\hat{\sigma}_X$ and electron diffraction $\hat{\sigma}_e$~\cite{Bartell64:JACS86}
\begin{eqnarray}
    \hat{\sigma}_X=\sum_i e^{i\vec{s}\cdot \vec{r}_i},\qquad\qquad\\
    \hat{\sigma}_e=\dfrac{1}{s^2}\left(\sum_{\nu}N_{\nu}e^{i\vec{s}\cdot\vec{R}_{\nu}}-\sum_{i}e^{i\vec{s}\cdot\vec{r}_i}\right),
\end{eqnarray}
where $\vec{s}$ is the momentum transfer vector and $N_{\nu}$ is the charge carried by $\nu$-th nuclear. The elastic and inelastic electron diffraction intensity for molecules in an electronic state $\ket{\psi_0}$ is~\cite{YangJ20:Science368}
\begin{eqnarray}
    I_e^{(\text{ela})}(\vec{s})&=&\frac{I_R}{s^4}\left|\sum_{\nu}N_{\nu}e^{i\vec{s}\cdot\vec{R}_{\nu}}-\sum_i\braopket{\psi_0}{e^{i\vec{s}\cdot\vec{r}_{i}}}{\psi_0}\right|^2\,,\\
    I_e^{(\text{inela})}(\vec{s})&=&\frac{I_R}{s^4}\left(\sum_{i,j}\braopket{\psi_0}{e^{i\vec{s}\cdot\vec{r}_{ij}}}{\psi_0}-\left|\sum_i\braopket{\psi_0}{e^{i\vec{s}\cdot\vec{r}_{i}}}{\psi_0}\right|^2\right)\,,
\end{eqnarray}
and for the x-ray diffraction intensity~\cite{Bartell64:JACS86}
\begin{eqnarray}
    I_X^{(\text{ela})}(\vec{s})&=&I_T\left|\sum_i\braopket{\psi_0}{e^{i\vec{s}\cdot\vec{r}_{i}}}{\psi_0}\right|^2\,,\\
    I_X^{(\text{inela})}(\vec{s})&=&I_T\left(\sum_{i,j}\braopket{\psi_0}{e^{i\vec{s}\cdot\vec{r}_{ij}}}{\psi_0}-\left|\sum_i\braopket{\psi_0}{e^{i\vec{s}\cdot\vec{r}_{i}}}{\psi_0}\right|^2\right)\,,
\end{eqnarray}
where $I_R$ and $I_T$ are Rutherford and Thomson scattering factors, respectively. 

\begin{figure}
    \centering
    \includegraphics[width=15cm]{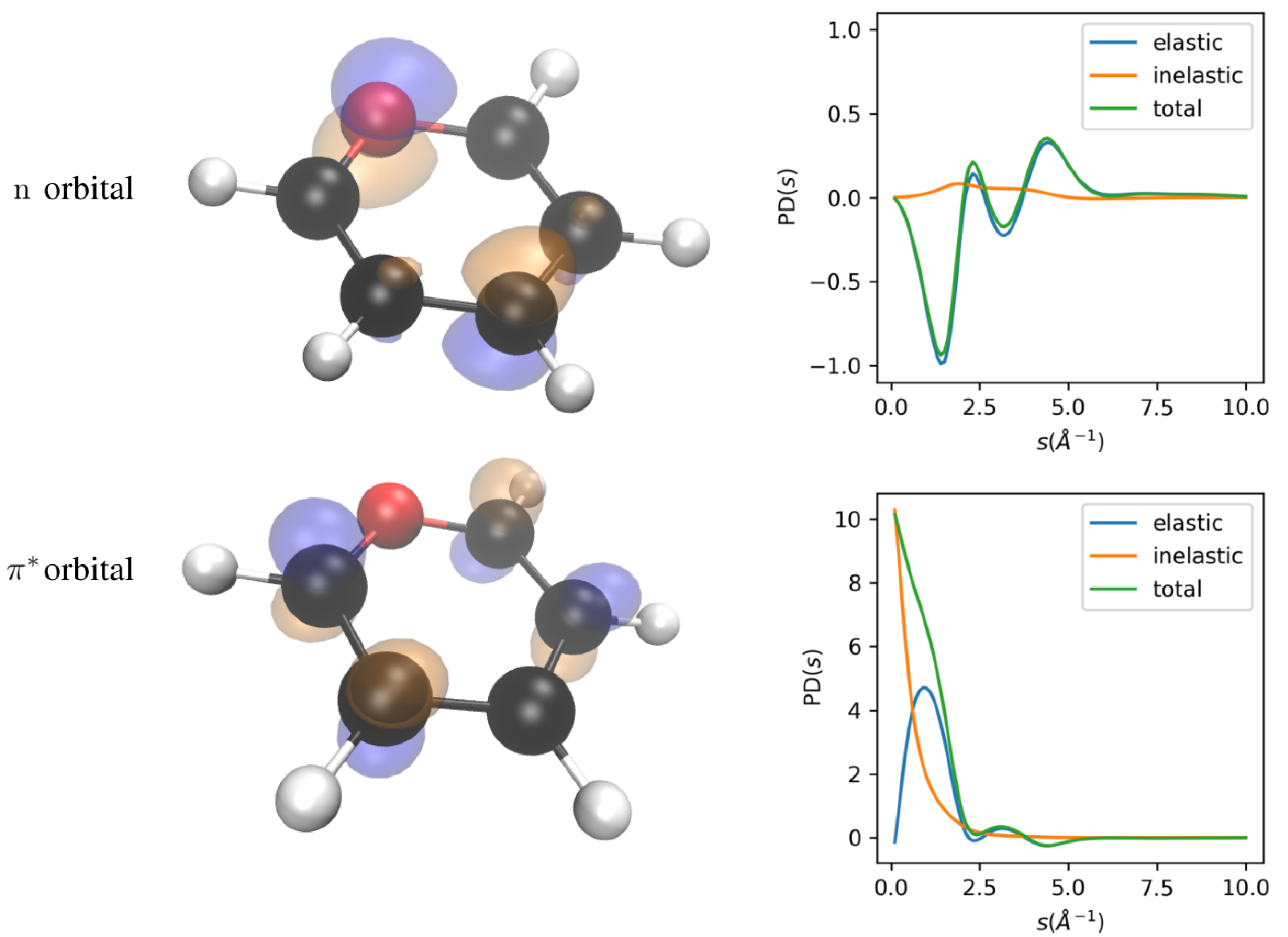}
    \caption{Comparison of percentage difference (PD) signals of x-ray and electron diffraction signal of pyridine in the $S_{\mathrm{0}}$ and $S_{\mathrm{1}}$(n$\pi^*$) states. Black, red and white spheres are C, N and H atoms, respectively. Isosurface representations of $\mathrm{n}$ and $\pi^*$ molecular orbitals are shown of value 0.1\AA$^{-3/2}$. The lone pair $\mathrm{n}$ orbital is localized at N and C atoms, where the electron correlation is much stronger than the anti-bonding $\pi^*$ orbital. As shown in the right column, the PD diffraction signal of x-ray diffraction (above) is smaller than electron diffraction (below) due to the difference of elastic signal background.}
    \label{fig:pyridine}
\end{figure}

The inelastic signal reflects electron correlation including exchange and Coulomb repulsion effects, which is related to two-electron reduced density $\rho^{(2)}(\vec{r}_1,\vec{r}_2)$ incorporating the probability of two-electron distance.
Thus for the small angle scattering, which is essential for electron correlation~\cite{YangJ20:Science368}, the additional denominator $s^4$ significantly enlarges the signal of smaller $s$ and improves the  accuracy of experimental measurements. Besides, the elastic background for electron diffraction is much smaller, for the two terms of $\hat{\sigma}_e$ contributing from electronic and nuclear parts cancel out at forward scattering direction, which eliminates the unchanged part of the signal and enhances the sensitivity of percentage difference signal. We simulated the percentage difference signal of x-ray and electron diffraction for the pyridine molecule $S_{\mathrm{0}}$ and $S_{\mathrm{1}}(\mathrm{n}\pi^{\star})$ state on the CASSCF(10,9)/cc-pVTZ level. As shown in Fig.~\ref{fig:pyridine}, the electron diffraction shows larger PD signals and could be more sensitive to the change of electronic correlation in electronic state transition.

{Gas-phase electron diffraction, as is discussed above in "Ultrafast X-ray Diffraction Imaging", does not show amplification of scattered radiation through Bragg reflections.
Pioneering works of Zewail and co-workers for the determination of transient molecular structures on picosecond timescale~\cite{Zewail03:HCA86} include non-equilibrium structures and radiationless processes. For example, in the nonconcerted elimination reaction of C$_2$F$_4$I$_2$, molecular structure of the transient intermediate (C$_2$F$_4$I) is determined as a nonbridged species~\cite{Hyotcherl01:Science291}. And dark structures in molecular radiationless transitions of several aromatic molecules can be directly determined using ultrafast electron diffraction~\cite{Remesh05:science307}.} The temporally evolving structure of photoexcited CF$_3$I molecule can be reconstructed by ultrafast gas-phase electron diffraction of laser aligned sample of isolated molecules~\cite{Hens12:PRL109}. The real space interatomic distances characterized by the pair distribution functions (PDF) can be extracted by Fourier transformation of diffraction pattern. Furthermore, the photodissociation and conical intersection dynamics of CF$_3$I has been directly imaged~\cite{YangJ18:Scicen361}.
{The coherent vibrational wavepacket of gas phase iodine molecules created by resonant laser excitation to $B$ state can be imaged by relativistic electrons diffraction. Beyond  nuclear position, the spreading of nuclear wave packets can also be captured~\cite{YangJ16:PRL117}. As another example, the rotational wavepacket of impulsively laser-aligned nitrogen molecule can be captured using tabletop keV-UED~\cite{Xiong20:PRR2} and MeV-UED~\cite{YangJ16:NC7} achieving similar temporal resolution of $\sim$200 fs, and one images the temporal evolution of N$_2$ rotational wavepacket and angular distribution changing from prolate to oblate. }

{Apart from the spatial charge effect and velocity mismatch problems, liquid-phase UED experiments have to tackle with the sample preparation procedure, since it is difficult to prevent liquid film from evaporation in vacuum without substrate. Recently an ultrathin liquid sheet jet enables MeV-UED by solving the preparation problem~\cite{Nunes20:SD7}. In liquid water, the ultrafast structural response to the excitation of the hydrogen bond (O-H) stretching vibration mode is measured by MeV-UED~\cite{YangJ21:Nature596}. A transient O-H bond contraction of roughly $0.04$~\AA~on a timescale of $80$~fs, followed by a thermalization on a timescale of approximately $1$~ps, is observed. Prior to the contraction of O-H bond, the difference pair distribution function ($\Delta$PDF) signal shows that the first intermolecular structural response is an O-O contraction, which provides a direct visualization of the intermolecular character of water vibration up to atomic resolution, and provide crucial evidence to resolve the long standing controversy concerning the relaxation pathway of vibrationally excited hydrogen bonding through Fermi resonant bending mode.
Another liquid-phase UED experiment looks into the ultrafast proton transfer in ionized liquid water~\cite{Lin21:Science374}, which produces hydroxyl radicals (OH) and hydronium cations ($\mathrm{H}_3 \mathrm{O}^+$). The short-lived radical-cation complex OH($\mathrm{H}_3 \mathrm{O}^+$) that was formed within $140$~fs through a direct
O-O bond contraction and proton transfer, followed by the radical-cation pair dissociation
and the subsequent structural relaxation of water within $250$~fs is captured by CPDF (charge-pair distribution function) and $\Delta$PDF signal in MeV-UED pattern.}

{The UED experiments of solid state systems have reached femtosecond temporal resolution much earlier than gas and liquid phase counterparts due to its uncomplicated preparation, and the thin solid film can avoid velocity mismatch with strong diffraction intensity. 
Thus keV-UED is already able to achieve the requirement for observing atomic motions in solids. The ultrafast nonthermal melting process in solid-liquid transition was observed by UED in aluminum~\cite{Siwick03:Science302}, bismuth\cite{Sciaini09:Nature458} and gold~\cite{Ernstorfer09:Science323}. 
By using short-pulsed lasers to deposit heat at a rate faster than the thermal expansion rate, visible and near-infrared laser light is absorbed by the electrons. 
In this way, intense short-pulsed optical excitation initially forms highly nonequilibrium conditions. UED for such solid state samples provides an atomic-level view of melting. Non-thermal melting for  aluminum and bismuth as well as the formation of warm dense matter in gold are explored with keV-UED patterns by analyzing intensities and pair correlation function (PCF). 
Another visualization for melting via UED happens on $30$-nm-thick polycrystalline tungsten films with highly populated defects caused by extreme radiation environment, which also confirms the molecular dynamic simulation for ultrafast melting~\cite{Mianzhen19:SA5}. 
Ultrafast electronic and lattice dynamics in materials undergoing charge density wave (CDW) phase transition, which are ideal model systems for the study of highly cooperative phenomena caused by interactions among electrons, phonons and spins on the nanoscale, are also observed by UED. 
In the quasi two-dimensional CDW system $1\mathrm{T}$-$\mathrm{TaS}_2$, optically induced change in the electronic spatial distribution drives atomic motions is observed, which are accompanied by a rapid electron–phonon energy transfer and are followed by fast recovery of the CDW~\cite{Eichberger10:Nature468}. 
And ultrafast photon-induced long-range $2$D ordered
electronic states at the surface in a $3$D CDW material $1\mathrm{T}$-$\mathrm{TiSe}_2$ is detected with MeV-UED, which is thought as a platform for realizing novel phases like superconducting state~\cite{Duan21:Nature595}.}

\textbf{Concluding Remarks.} We have presented the current state of the art in imaging molecular dynamics with tabletop low-frequency ultrafast laser, x-ray free electron lasers and electron pulses. 
The imaging by laser induced photoionization and electron rescattering, and by ultrafast x-ray and electron diffraction are versatile and powerful tools for mapping dynamic molecular structure with numerous advantages. {For example, interior structural detection of materials is possible due to small absorption probability and long penetration depth of x-ray~\cite{Santra09:JPB42}. Coulomb scattering provides large cross section for electron diffraction which shortens the exposure time~\cite{Zewail03:HCA86}.} 
In comparison with spectroscopic methods, where the structural information is indirectly inferred based on exact knowledge of energy landscape, ultrafast imaging techniques are more effective to provide direct access to atom distances and intuitive view of chemical reactions~\cite{YangJ16:NC7}. {For example, laser spectroscopy can give ample evidence for radiationless transitions, but the information of nuclear motion driving electronic transition can not be provided~\cite{Domcke20:Sci368}. Diffraction imaging unequivocally reveals electronic and nuclear dynamics~\cite{YangJ20:Science368}, which resolves a decades-old puzzle in molecular spectroscopy~\cite{Domcke20:Sci368}.} However, to image complex molecule remains challenging for the indistinguishable overlap of the pair distribution functions from different atom pairs in the diffraction imaging. Instead of comparing and optimizing theoretical model to best fit the data, {it would be more powerful to direct retrieval of structural information~\cite{Martin16:JPB49}. X-ray and electron scattering experiments can obtain the position of each atom within a molecule~\cite{YangJ16:NC7}, and in principle, offer direct access to complete molecular structures~\cite{Ma20:Struct7}. However, direct inversion from experimental data to quantum state of system is challenging for general systems. Even for classical ball-and-stick model, recovering structural parameters like bond lengths and bond angles for complex molecules is still difficult. The existence of inversion algorithm and its accuracy is often limited by the dimension of measurement data and experimental resolution.} Further development including brighter and shorter light or electron sources, higher spatio-temporal resolution and more robust structure retrieval methods will be needed for making molecular movies, especially for complex molecules.

\section*{Acknowledgement}
We thank the support of National Natural Science Foundation of China (Grant Nos. 92050201, 12174009, 11774013).
The authors thank Gopal Dixit for helpful discussions.

\end{document}